# New approach for building of unified theory about the Universe and some results

S. Sarg

E-mail: sarg@helical-structures.org

Web site: www.helical-structures.org

The physical models of a successful unified theory about the Universe must operate in different phase of matter evolution and different fields of physics. The attempts to build such wide range theory as a bunch of theories developed for different fields of physics are not quite successful. The accumulated knowledge from experiments and observations leads to a conclusion that some of the adopted postulates in the modern physics are not absolutely fundamental, as considered so far. A new approach for building of unified model of the Universe suggests resurrection of the principles of causality and logical understanding for any kind of physical phenomena. It is successfully applied in a new theory titled Basic Structures of Matter, which provides fundamentals for a unified theory about the Universe. The new approach leads to different physical models for the elementary particles and the atoms and also to a different concept about the Universe. In the same time the suggested models exhibit the same interaction energies as obtained by the Quantum mechanics and experiments. The analysis of the physical phenomena from a new point of view allows deeper understanding of the relations between the basic physical attributes: mass, energy, space, time, gravitation and inertia.

## 1. Problems related to development of successful unified theory about the Universe.

The foundations of the modern physics rely on postulates and rules adopted about 100 years ago. Now a real gap exists between the highly abstractive theories about the matter from one side, and physical reality from the other. This gap has been in favor of separation of the theoretical physics as an independent field. In the beginning of the 20th century the physical science has not been able to provide logical explanations of some physical phenomena, while their mathematical interpretations appeared successful. As a result the Quantum Mechanics is born. The principle of causality, that is essentially important for the real objectivity, has been replaced by the uncertainty principle. Influenced by the successful results from the quantum mechanical models the theoretical physics gradually abandoned the physical logic. It was the mathematical logic that took the guiding role in the analysis of the physical phenomena. The mathematical physics now dominates in the theoretical vision about matter, space and time. This approach led to development of number of highly abstractive theories in both directions of length scale reference to the Quantum mechanics range: into micro and macro cosmos. In the same time attempts are made to interconnect or extend the ranges of the existing theories into one universal unified field theory, valid for wide range of space and time. Many of the rules adopted in the separate narrow range theories, however, appear cross violated in the wider range of space and time. Such approach evidently is not quite successful for building of unified theory about the Universe.

The most convincing achievements of the Quantum mechanics are related with the atomic and molecular spectra, because they are validated by the experimental data. The range of these interactions, however, occupy a small range of the space and time scales in comparison to the vast space and time scales from micro to macro Cosmos. The validation of developed models for interactions away from this range is not so effective. In the bottom range of the scale the time events are very fast and the observations are characterized with large statistics but deteriorated time and space resolution of the interaction event. This is valid for particle physics experiments. In the upper scale, the space resolution is good but the observations lack the track of event time development. This is valid for cosmological phenomena. Presently, large number of physical phenomena predominantly in the range of micro and a macro cosmos are not satisfactorily explained.

In the range of macro cosmos: The dipole shape of the Cosmic Background Radiation, the deviation of the Hubble law from linearity for redshift $z > 0.8$, the Lyman alpha forest, the physics of Globular Clusters and the cepheids of second type, the red shift periodicity in the spectrum of quasar objects, the large velocities of most of the pulsars and almost zero velocities of others, the galactic rotational curves and a dark matter concept, the large energy in the galactic center, a black hole emitting radiation (recently discovered), the pillars formation (like in Crab nebulae) and their synchrotron type of radiation, the God's effect (the intergalactic distances are increasing in any radial direction from us), the gamma ray burst phenomena and so on.

In the range of micro cosmos: a single photon interference and polarization, the anomalous magnetic moment of the electron, the infinities in the Faynman diagrams, the large masses of bosons and the top quark (larger than the mass of the atom of gold), CPT theorem and its violation, parity violation in Beta decay of charge particles, impossibility to create a stable antihydrogen atom for not relativistic velocities.

The Quantum mechanics is undoubtly successful theory. At the same time it is so bizarre, from the point of view of human logic, that Richard Feynman, a master of quantum calculation said: "no-body understand the quantum mechanics" (how it works). Many rules adopted in the quantum mechanics are not supported by logical understanding. Then a question arises: Do the quantum mechanical models reflect real physical entities or they are pure mathematical models? If the latter option is the case they could not be logically extended into both directions: the micro and the macro cosmos.

## 2. The absolute validity of some of the adopted postulates is challenged by the accumulated knowledge.

In recent years number of publications challenging some of the adopted postulates of the modern physics appeared in the peer review journals. They touch fundamentally important physical attributes: the gravitation, the zero point energy, the vacuum properties and the inertia.

One of the discussed problems is the controversy over the Newton's gravitational constant. Despite it has been measured since 1798 (initially by Cavendish) its currently known value is much less accurate than many physical constants. This problem is investigated in many universities and institutions over the world. The new results measured by different teams also disagree widely. One of the team providing well informative website is the Eot-Wash group in the University of Washington, Seattle (http://www.npl.washington.edu/etwash). Another international group of investigators reports quite interesting experimental results in the article "Experimental evidence that the gravitational constant varies with orientations", by M. L. Gershteyn et al. [1]. In this paper the researchers found that the gravitational constant varies significantly with the orientation of the masses relative to the system of fixed stars.

Another discussed topic, related to gravitation, is the understanding of the vacuum and its properties. The potential benefits could be a new revolutional discovery. The paper "Can the Vacuum be Engineered for Spaceflight applications?, by H. E. Puthoff [2] has been reported at the **NASA Breakthrough Propulsion Physics conference**, 18 Aug, 1997. The following paragraphs are extracted from the paper:

“With the rise of special relativity which did not require reference to such an underlying substrate, Einstein in 1905 effectively banished the ether in favor of the concept that empty space constitutes a true void. Ten years later, however, Einstein’s own development of the general theory of relativity with its concept of curved space and distorted geometry forced him to reverse his stand and opt for a richly-endowed plenum, under the new label spacetime metric.”

"After a one-year investigation Forward finished his study and submitted his report to the Air Force, who published it under the title Mass Modification Experiment Definition Study. The Abstract reads in part: "...Many researchers see the vacuum as a central ingredient of 21st Century physics. Some even believe the vacuum may harness to provide a limitless supply of energy. .... **It was possible to find an experiment that might be able to prove or disprove that the inertial mass of a body can be altered by making changes in the vacuum surrounding the body**."

Another article of H.E. Puthoff et al. in the First International Workshop in Field Propulsion is titled: “Engineering the Zero-Point Field and Polarizable Vacuum for Interstellar Flight [3].

## 2. Advantages of the suggested new approach

The presented historical overview and the emphasized problems are in agreement with the considerations expressed by number of physicists, that the principles of causality, objective reality and logical understanding must be resurrected. This means that the physical science must be put on logical rails by development of physical models understandable by the human logic and reflecting the reality. In the same time the useful achievements of the modern physics should not be undermined. The working mathematical models could be used in a way that their results could be filtered by the understandable physical models. This could be a kind of a new approach in the contemporary physics. Using this approach a successful unified theory could be built by testing the suggested new physical models in different fields of the physics. The analytical algorithm of this approach will be distinguished from the existing current approach by the following:

- The suggested physical models and basic laws must allow analysis at lowest level of matter organization in 3 dimensional space and unidirectional time. The space and time parameters at this level must be separated.
- The preserved physical logic and understanding will allow cross analysis between different fields of the physics

- A trend in matter evolution in cosmological scale could be identified

- The analytical process may lead to closed cycles. In such case an iteration kind of analysis could be applied and some hidden parameters and phenomena could be unveiled.

The proposed new approach is successfully applied in the thesis Basic Structures of Matter [4], presenting an original concept of unified theory about the Universe.

## 4. Concept of BSM theory and conclusions obtained by analysis of experimental data and observations.

The suggested concept allows to obtain logical explanations of all kind of phenomena from the micro to macro Cosmos using a three-dimensional space and unidirectional time scale. In the same time it allows mathematical treatment of the investigated phenomena in order to express the unveiled new physical parameters by the known physical constants. The provided analysis of phenomena from different fields of physics leads to the following conclusions:

- The energy is inseparable attribute of the matter.

- The vacuum is not a void space, but contains a unique grid structure. This grid structure named a Cosmic Lattice (CL) is built by two types of alternatively arranged nodes, each one containing 4 sub-elementary particles with shape of six sided prisms. These two particles are formed of two not mixable types of intrinsic matter substances with different density ratio. They are distinctive also by their length ratio (3:2) and by the twisting feature of their internal structure - left and right handed. Their estimated lengths are in the range of (1 to 10) x $10^{-20}$ (m).
- Prisms of same matter substance are attracted in empty space by Intrinsic Gravitational (IG) forces inverse proportional to the cube of the distance.
- The CL node possesses energy well and freedom for complex spatial oscillations characterized by two kind of proper frequencies. Spontaneously synchronized CL nodes forms domains with zero point waves equalizing the zero point energy of the vacuum. The dynamical parameters of the CL nodes included in such domains are related to the physical parameters permeability and permittivity of the free space.
- The structure named a Cosmic Lattice is spread in the visible Universe. So the observable CL space possesses quantum features and creates conditions for fields: gravitational, electric and magnetic.
- The quantum features of CL space are responsible for the constant light velocity.
- The photon is an energy-propagating wave with specific spatial structure, boundary features and quantum features provided by the CL space environments.
- The Newton's gravitation (universal gravitational law) is a propagation of the IG field in CL space.
- The atomic particles are complex formation of helical structures made of same type prisms held by IG forces and possessing denser internal lattice structures. They are formed in unique process of crystallization in a hidden phase of a home galaxy evolution.
- The CL space have three important parameters: Static CL pressure (related with the mass we are familiar with, Dynamic CL pressure (related with the Zero Point Energy of the CL space (vacuum) an Partial CL pressure (related with inertial properties of the atomic matter in CL space). The observable temperature of 2.72 K from the free space is a signature of the Zero Point Energy of the CL space.
- Every massive object possesses own CL space extended beyond its surface and defining the local conditions of an inertial frame (Special Relativity) and a space curvature (General Relativity. Separation zone between two massive objects exists providing boundary of the local CL spaces. The extension of the local CL space is subordinated by the inverse cubic IG law.
- The flexible geometry of the CL node allows folding and unfolding with a speed rate of light velocity. Moving of lighter object in CL space of heavier one is related with continuous process of folding and unfolding of CL nodes. This kind of interaction is included in the inertia of the atomic matter (property we are familiar with). The energy momentum carried by the folded nodes is related to the parameter Partial pressure of CL space. The following derived relation is valid for the CL spaces of all connected galaxies in the Universe:

$$P_P / P_S = \alpha^2 / \sqrt{1-\alpha^2} \qquad \text{[Eq. (10.18) of BSM]}$$

where: $P_P$ and $P_S$ - are partial and static CL space pressure, respectively and α - is a fine structure constant.

- Every galaxy undergoes through a cosmological galaxy cycle containing a phase of active life and hidden phases. In the hidden phases the galactic matter is separated from the neighboring galactic CL spaces so it is invisible. In this phases the atomic matter is recycled to the level lower than the prisms. The Gamma Ray Burst phenomena are related with a collapse or a birth of a galaxy in a same space of the Universe.
- The fine structure constant originates in the lowest level of matter organization, but its signature is propagated up to the atomic and molecular levels. Its signature is also involved in the inertial properties of the atomic matter in CL space environments.
- The material origin of the vacuum structure and the complex helical structures of the elementary particles have not been taken into account in the existing models of the scattering experiments. The result is a significant deviation of the calculated dimensions of the stable atomic particles from their real physical dimensions.
- The Bohr atomic model is a correct mathematical model when assuming that the space is void. The real physical model in structured vacuum, however, is completely different.
- In the BSM concept the Quantum mechanical rules are derivable and the effects of the Special and General relativity are understandable.
- The Universe is stationary and not homogeneous. The observed red shift of the galaxies is not from a Doppler shift, but from accumulation effect of energy losses when the photons pass the boundaries between the different galaxy CL spaces. The galaxy CL space differences come from small differences in geometrical parameters of molded prisms. The molding process takes place in one of the hidden phases of the galaxy evolution. In this process the total amount of intrinsic matter and energy of the galaxy is involved. The prism matter quantity, however, appears always the same due to a large histeresis in the intrinsic matter-energy interaction process.
- The Cosmic Microwave Background radiation (CMB) is a signature of the Zero Point Energy (ZPE) possessed by the CL space, while the experimentally estimated temperature of 2.72 K is a CL space background temperature. The spatial (dipole) anisotropy is a result of inertial interactions between the atomic matter of our local frame (local CL space) and the home galaxy CL space. In Chapter 5 of BSM thesis a theoretical expression for the background temperature is derived, based on the dynamic CL pressure (carried by the Zero Point Waves) on the protons of the hydrogen atoms or molecules.

$$T = \frac{{N_A}^2 h \nu_c {L_{PC}}^2 (R_C + r_p)^3}{2 r_e R_c c S_W R} \frac{\mu_e}{\mu_n} = 2.676 \text{ K} \qquad \text{[Eq. (5.8) of BSM]}$$

where: $N_A$ – Avogadro constant, $h$ – Planck constant, $\nu_c$ - Compton frequency, $L_{PC}$ – proton core length ($L_{PC}$ = 1.6277 Angstroms, cross-calculated) , $R_C$ – Compton radius of electron, $r_e$ = 8.84E-15 (m), $r_e$ and $r_p$ – small radii of electron structure derived by cross analysis and related by a ratio $r_e / r_p = 3/2$ (same as prisms length ratio), $S_W$ = 1 $m^2$ – reference surface in SI system, R – ideal gas constant, $(\mu_e / \mu_n)$ - magnetic moment ratio between electron and neutron (magnetic moment for the neutron is used instead of for proton in order to reflect the neutrality of the atom or molecule), c – light velocity used as a factor (for zero point wave interactions)

- The mass we are familiar with is an attribute of the atomic level of matter organization in CL space environments. The equation $E = mc^2$ reflects the energy obtained from destruction or "hiding" of the atomic matter particles. The intrinsic matter could never disappear.
- The intrinsic gravitation is a kind of energy balance originated in the lowest level of matter organization.

Fig. 1 illustrates the position of known and unveiled structures in the length scale of the Universe. Fig. 2 shows the period of unveiled cycles related to these structures in the time scale of the Universe.

Fig. 3 illustrates the difference between the new analytical approach applied by the BSM theory and the existing one. The broken gray line in the existing so far approach indicates the fields oh physics where the logical understanding fails. The BSM approach is distinguished by two principal features: (a) the logical understanding is fully preserved; (b) a repeatable cycle of the matter evolution is identified. The latter feature is presented in Fig. 3 as a loop. It allows unveiling of some hidden physical phenomena by using an iterative method of logical analysis.

The proposed BSM concept is verified by cross validation of the output results of the suggested models with experimental data from different fields of the physics. The unveiled low-level structures of the vacuum and the elementary particles further allowed deciphering the structures of the proton and neutron and their spatial arrangement in the atomic nuclei, as well. It appears that the row and column pattern of the Periodic table carry a signature of the nuclear structure. The electron orbitals are strictly defined by the protons and neutrons arrangement in the nucleus. The features of the Hund's rule and Pauli exclusion principle are also identifiable. So one of the output results of the BSM is an illustrative appendix titled Atlas of Atomic Nuclear Structures (ANS) [5]. It provides the arrangement of the protons and neutrons in the atomic nuclei for elements from Hydrogen to Lawrencium (Z=103). Fig. 4 shows views of the physical models of some selected elements.

The success of the BSM theory for explanation of large diversity of physical phenomena is a result of the applied new approach.

## 5. Advantages and potential applications of BSM theory.

One of the main advantages of the BSM theory is the provided logical understanding. All suggested models are real physical entity operating in three dimensional space. The interaction processes are in unidirectional time scale, so the principle of causality is preserved. These features greatly facilitate the analysis of complex physical phenomena. In this category are included: the effects of the Special and General Relativity, the Zero Point Energy of the vacuum, the superconductivity, the ferro and dia magnetism, the gravity and the inertia. In chapter 10 of BSM the inertial properties of the atomic matter in CL space are discussed. The Newton's law about inertia (F=ma) is understandable interaction process. Relativistic considerations for interstellar space travels are discussed. One of the potential applications is presented in a separate article titled "New vision about a controllable fusion reaction with efficient energy yield" [6].

**References**:

[1] Mikhail L. Gershteyn et al., Experimental evidence that the gravitational constant varies with orientation, (http://arxiv.org/abs/physics/0202058).

[2]. H. E. Puthoff, Can the Vacuum be Engineered for Spaceflight applications?, NASA Breakthrough Propulsion Physics conference at Lewis Res. Center, Aug 1997.

[3]. H. E. Puthoff, S. R. Little, M. Ibison, Engineering the Zero-Point Field and Polarizable Vacuum for Interstellar Flight, First International Workshop in Field Propulsion, University of Sussex, Brighton, UK, Jan 2001.

[4]. S. Sarg, Basic Structures of Matter (thesis on matter, space and time), www.helical-structures.org/, also in the National Library of Canada:

http://amicus.nlc-bnc.ca/aaweb/amilogine.htm (AMICUS Web search by title).

[5] S. Sarg, Atlas of Atomic Nuclear Structures,

www.helical-structures.org/, (also AMICUS Web search by title)

[6] S. Sarg, New vision about a controllable fusion reaction with efficient energy yield, www.helical-structures.org/Applications file bsm_application_1.pdf.

Note: The articles [2] and [3] are available in the Web site: www.nidsci.org/articles/articles3.html

Used abbreviations: BSM – Basic Structures of Matter (theory); CL – Cosmic Lattice, IG – Intrinsic Gravitation.

## Figures

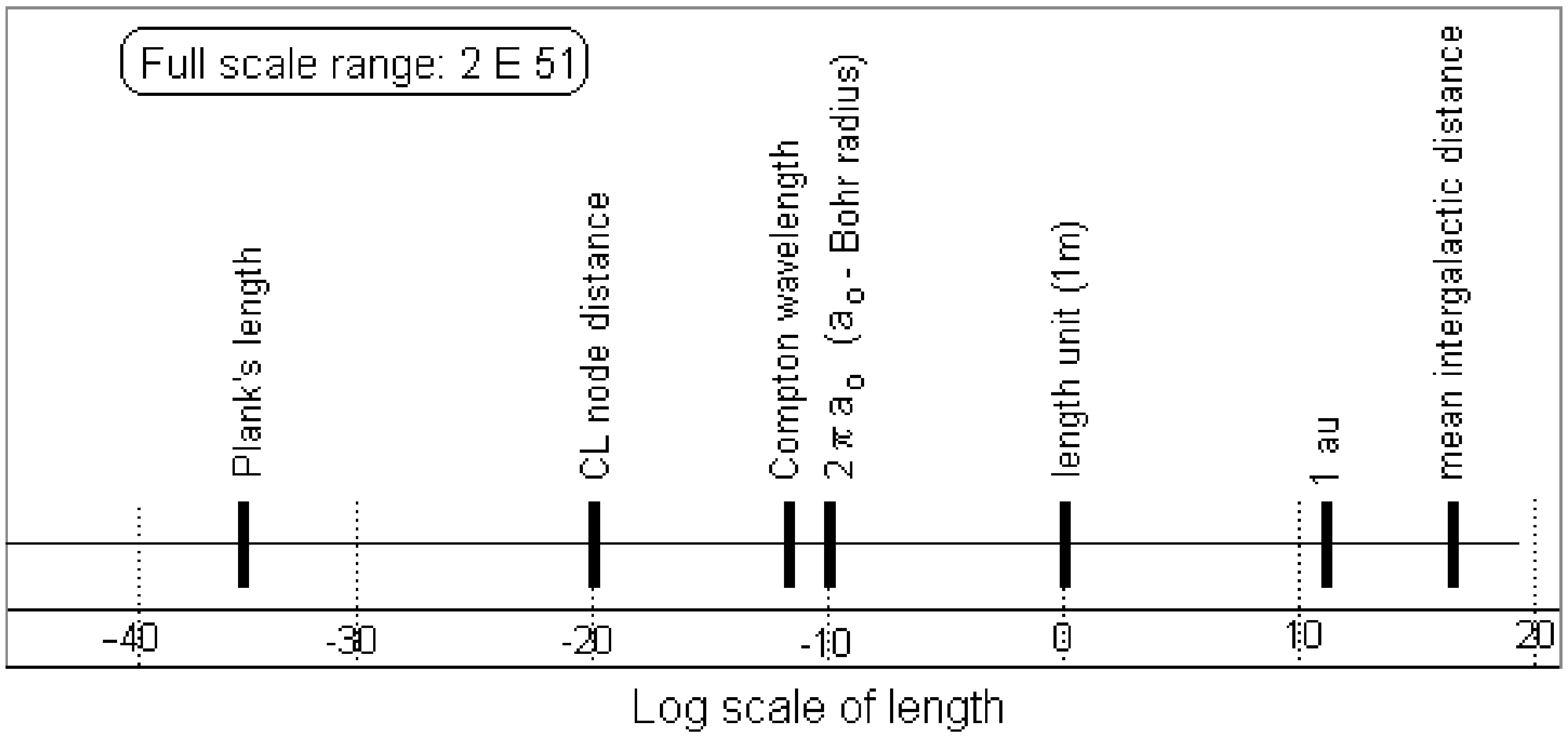

Fig. 1

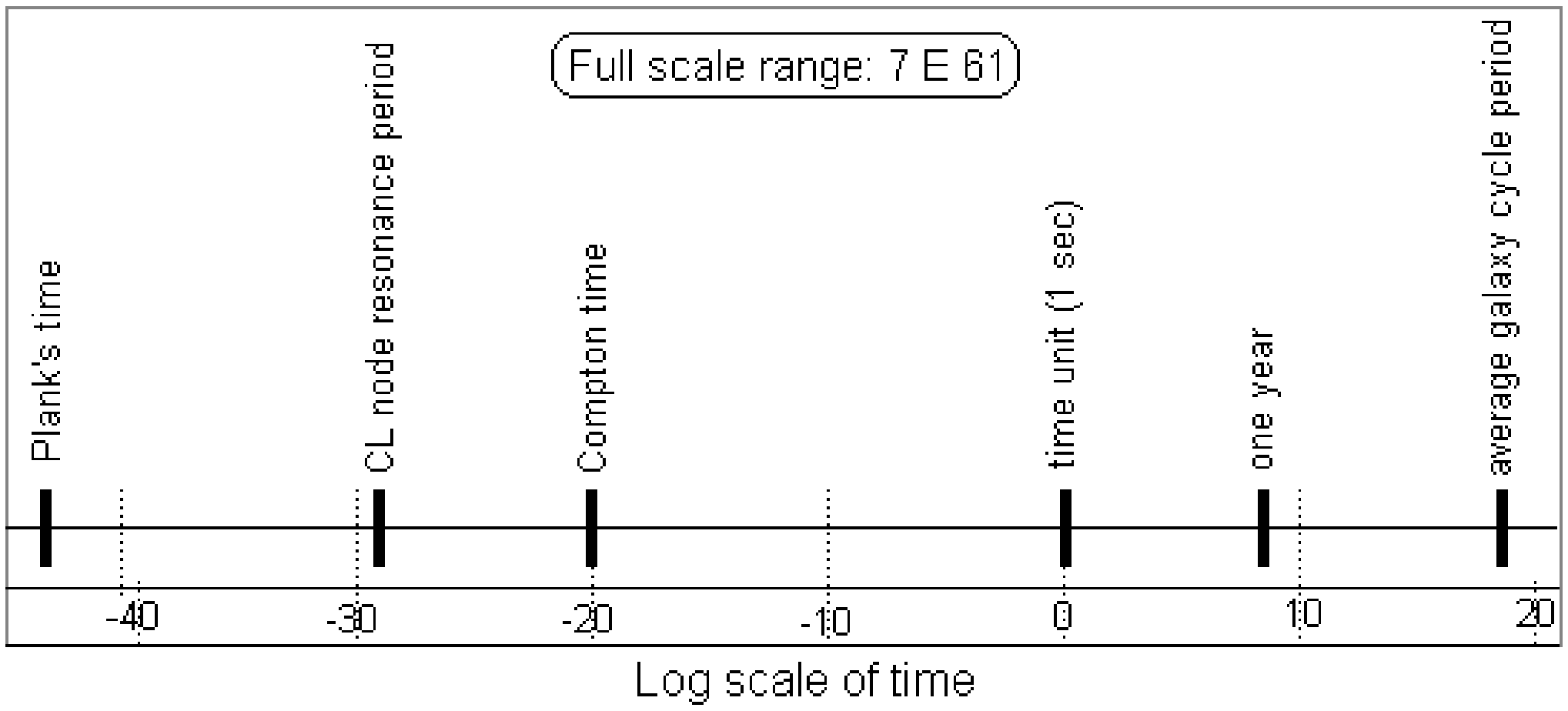

Fig. 2

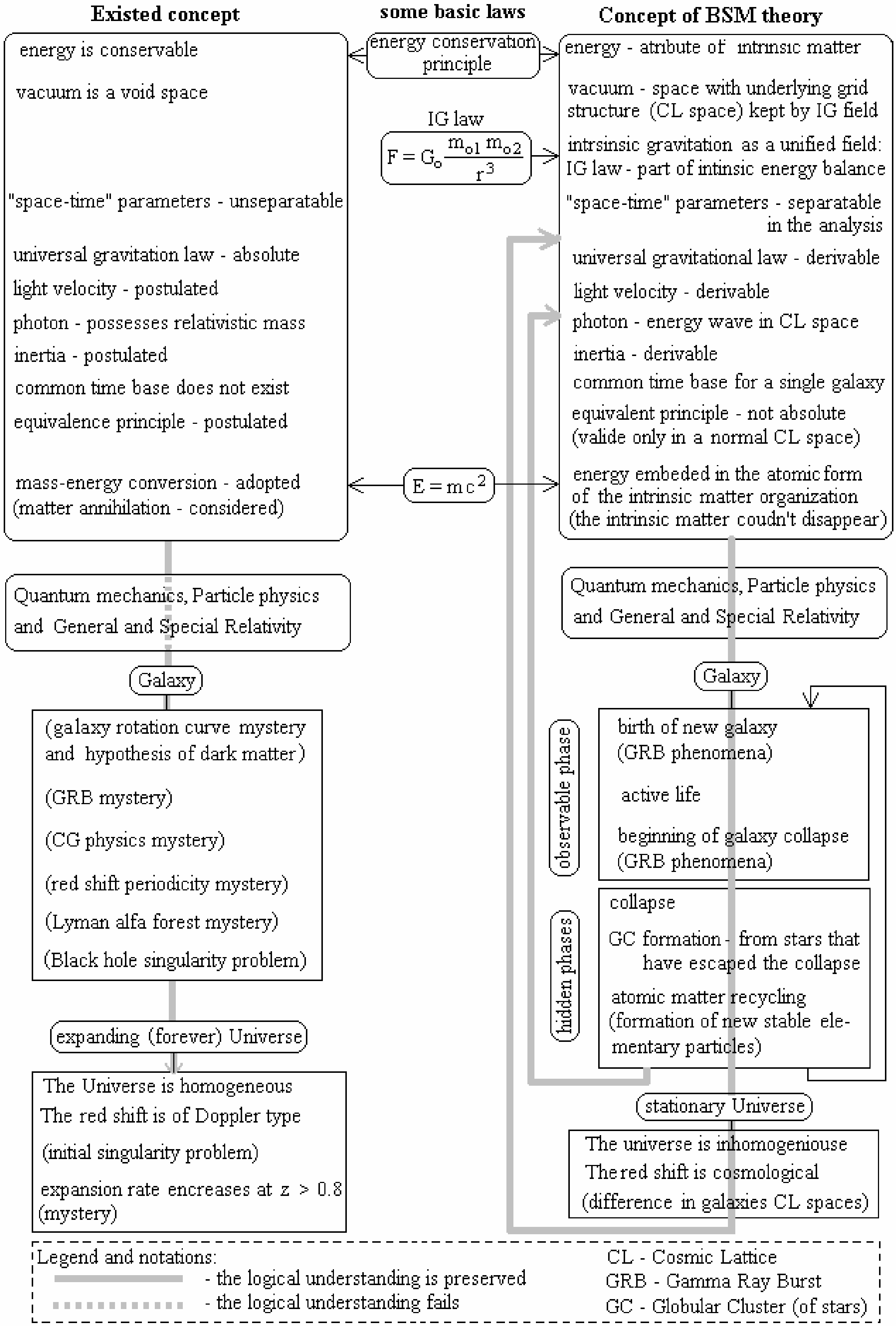
Existed concept
some basic laws
Concept of BSM theory
energy is conservable
vacuum is a void space
"space-time" parameters - unseparatable
universal gravitation law - absolute
light velocity - postulated
photon - possesses relativistic mass
inertia - postulated
common time base does not exist
equivalence principle - postulated
mass-energy conversion - adopted
(matter annihilation - considered)
energy conservation principle
IG law
$F = G_o \frac{m_{o1} m_{o2}}{r^3}$
$E = mc^2$
energy - atribute of intrinsic matter
vacuum - space with underlying grid structure (CL space) kept by IG field
intrsinsic gravitation as a unified field: IG law - part of intinsic energy balance
"space-time" parameters - separatable in the analysis
universal gravitational law - derivable
light velocity - derivable
photon - energy wave in CL space
inertia - derivable
common time base for a single galaxy
equivalent principle - not absolute (valide only in a normal CL space)
energy embeded in the atomic form of the intrinsic matter organization (the intrinsic matter coudn't disappear)
Quantum mechanics, Particle physics and General and Special Relativity
Galaxy
(galaxy rotation curve mystery and hypothesis of dark matter)
(GRB mystery)
(CG physics mystery)
(red shift periodicity mystery)
(Lyman alfa forest mystery)
(Black hole singularity problem)
expanding (forever) Universe
The Universe is homogeneous
The red shift is of Doppler type
(initial singularity problem)
expansion rate encreases at z > 0.8 (mystery)
Quantum mechanics, Particle physics and General and Special Relativity
Galaxy
observable phase
birth of new galaxy (GRB phenomena)
active life
beginning of galaxy collapse (GRB phenomena)
hidden phases
collapse
GC formation - from stars that have escaped the collapse
atomic matter recycling (formation of new stable elementary particles)
stationary Universe
The universe is inhomogeniouse
The red shift is cosmological
(difference in galaxies CL spaces)
Legend and notations:
- the logical understanding is preserved
- the logical understanding fails
CL - Cosmic Lattice
GRB - Gamma Ray Burst
GC - Globular Cluster (of stars)

**Fig. 3**

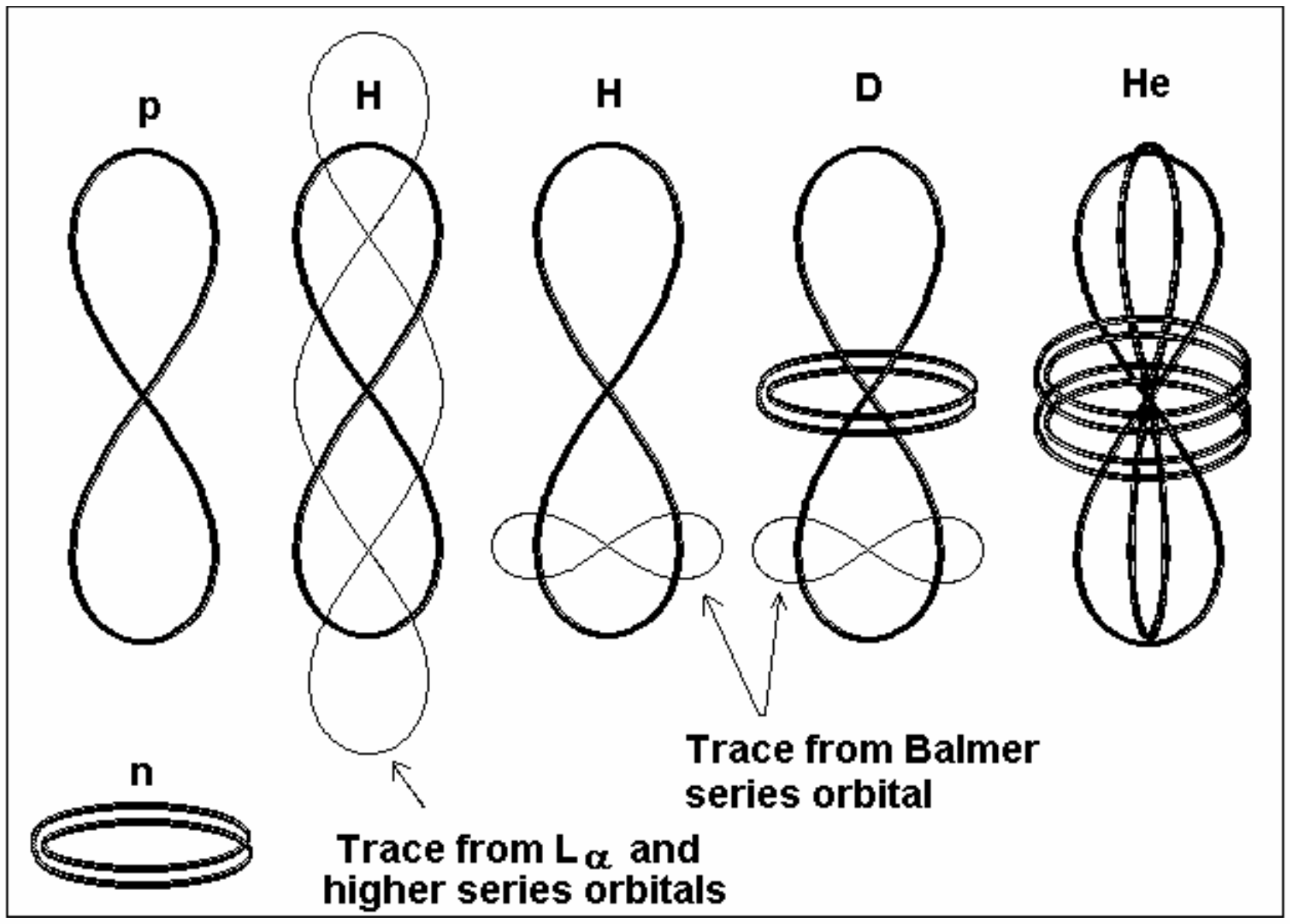

Fig. 4

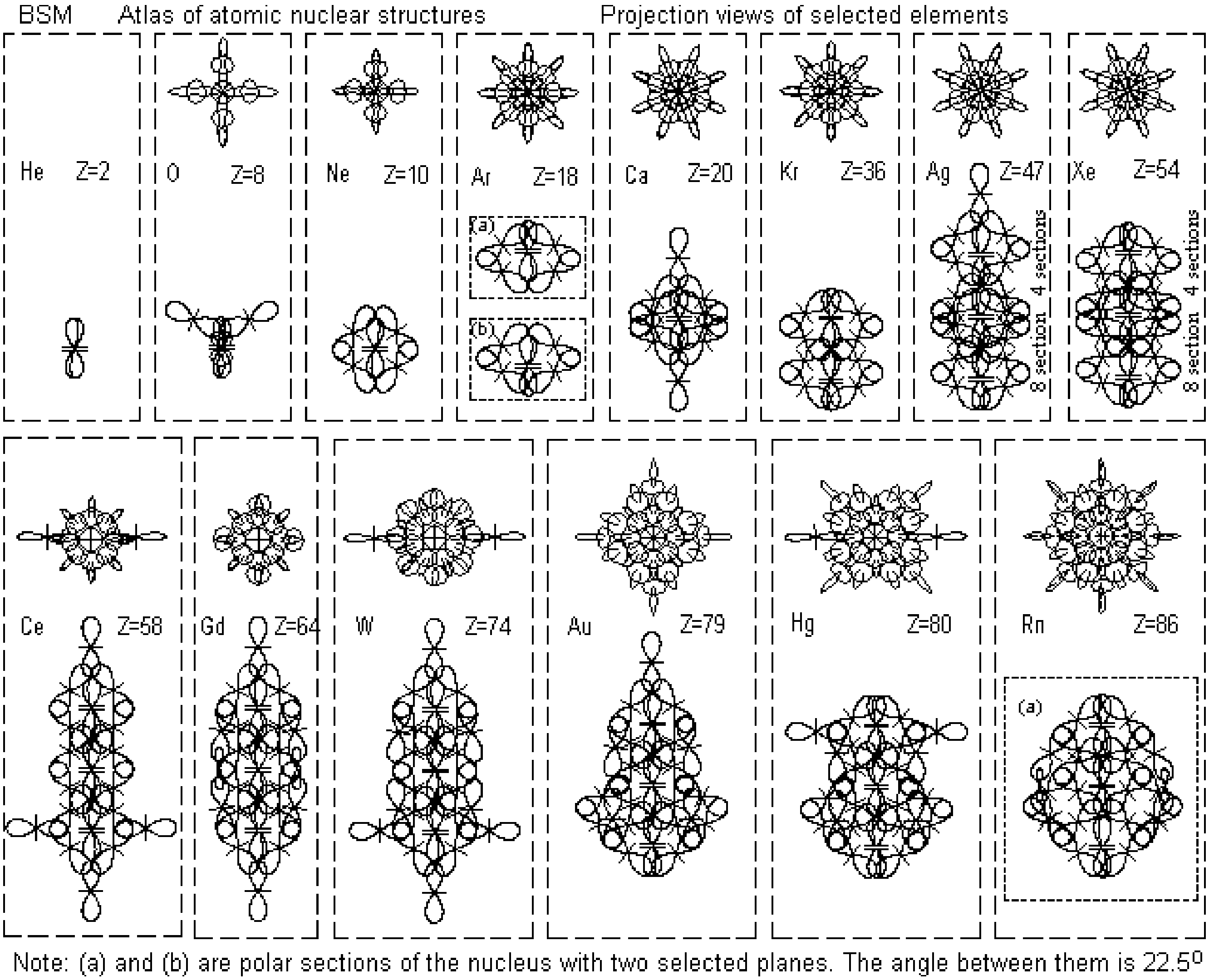

**Fig. 5**. Selected elements from the Atlas of Atomic Nuclear Structures